# Cyber-infrastructure to Support Science and Data Management for the Dark Energy Survey

C. Ngeow[*a], J.J. Mohr[a,b], T. Alam[b], W.A. Barkhouse[a], C. Beldica[b], D. Cai[b], G. Daues[b], and R. Plante[b]

[a]University of Illinois Department of Astronomy, 1002 W. Green St, Urbana, IL USA 61801
[b]National Center for Supercomputing Applications, 1205 W. Clark St., Urbana, IL USA 61801

J. Annis[c], H. Lin[c], and D. Tucker[c]
[c]Fermi National Accelerator Laboratory, P. O. Box 500, Batavia, IL USA 60510

R.C. Smith[d]
[d]Cerro Tololo Inter-American Observatory/NOAO, 950 N. Cherry Ave, AZ USA 85719

**ABSTRACT**

The Dark Energy Survey (DES; operations 2009-2015) will address the nature of dark energy using four independent and complementary techniques: (1) a galaxy cluster survey over 4000 deg$^2$ in collaboration with the South Pole Telescope Sunyaev-Zel'dovich effect mapping experiment, (2) a cosmic shear measurement over 5000 deg$^2$, (3) a galaxy angular clustering measurement within redshift shells to redshift=1.35, and (4) distance measurements to 1900 supernovae Ia. The DES will produce 200 TB of raw data in four bands, These data will be processed into science ready images and catalogs and co-added into deeper, higher quality images and catalogs. In total, the DES dataset will exceed 1 PB, including a 100 TB catalog database that will serve as a key science analysis tool for the astronomy/cosmology community. The data rate, volume, and duration of the survey require a new type of data management (DM) system that (1) offers a high degree of automation and robustness and (2) leverages the existing high performance computing infrastructure to meet the project's DM targets. The DES DM system consists of (1) a grid-enabled, flexible and scalable middleware developed at NCSA for the broader scientific community, (2) astronomy modules that build upon community software, and (3) a DES archive to support automated processing and to serve DES catalogs and images to the collaboration and the public. In the recent DES Data Challenge 1 we deployed and tested the first version of the DES DM system, successfully reducing 700 GB of raw simulated images into 5 TB of reduced data products and cataloguing 50 million objects with calibrated astrometry and photometry.

## 1. INTRODUCTION

The Dark Energy Survey (DES) is a project to study the dark energy or expansion history of the universe that requires an extended multi-band survey. It is designed to include scientific redundancy to help control systematic effects that can bias individual analyses. We will study the dark energy through four independent techniques as described below. Each of these techniques could stand alone as a probe of the dark energy, and each study will be by far the best in its class when the survey ends. An important strength of our approach is that these measurements are subject to different systematic uncertainties. We are designing each study to minimize these systematic uncertainties, and by simply comparing independent constraints we will obtain important crosschecks. Because each of these studies has different observational leverage on the dark energy and other cosmological information, the combined constraints from the four measurements will push well beyond what each can do. With appropriate focus on controlling systematics, it is in principle possible to use our study to differentiate dark energy models from models that alter the behavior of gravity. These four techniques[1] are:

1. *A galaxy cluster survey extending to a redshift ~1.35 and detecting ~30,000 clusters*. In coordination with the South Pole Telescope Sunyaev-Zel'dovich effect mapping experiment (that will operate at mm wavelengths), the galaxy cluster survey over 4000 deg$^2$ will constrain the energy densities and natures of the components of the universe through measurement of the redshift distribution of galaxy clusters and through the spatial clustering of the galaxy clusters. Essentially, such a survey produces a census of the abundance of galaxy clusters from the present time to a time when the universe was only a third its present age. Changes in cluster abundance tell us how rapidly clusters have formed in the past, and the rate of formation is quite sensitive to the nature of the dark energy[2,3].

---
[*] Contact email: cngeow@uiuc.edu

2. *A weak lensing study of the cosmic shear extending to large angular scales.* Weak lensing can be used to study the cosmic shear, which is a measure of the gravitational lensing effect of the matter distribution in the nearby universe on distant galaxies. Cosmic shear measurements constrain the nearby dark matter distribution in a very direct way, and the properties of the dark matter distribution contain a wealth of information about the components of the universe, including the nature and density of the dark energy.

3. *A galaxy angular power spectrum study within redshift shells to redshift ~1.35 using 300 million galaxies.* This is a measure of the clustering in the galaxy population within redshift shells extending from the present to a time when the universe was only a third its present age. The angular power spectrum of galaxies reflects the underlying clustering of the dark matter distribution in those same redshift shells and exhibits additional features called the baryonic acoustic oscillations. With this technique we can map out how this dark matter distribution is changing with time and infer the mean distance to each redshift shell[4], providing constraints on the dark energy.

4. *A time domain study that will deliver ~1900 supernovae Ia distance estimates.* This technique was used to discover the dark energy in 1998[5,6], and it remains as one of the key tools for improving our understanding of the dark energy. We will devote 10% of the survey time to repeated observations of ~40 deg$^2$ area, which will let us find supernovae and then follow the evolution of the brightness of these supernovae. This will yield ~1900 supernova Ia distance measurements between redshift 0.3 and 0.8, and it will make the DES the largest supernova study of dark energy until the next generation supernovae surveys begin their science operations.

Achieving the DES science goals requires building two tools: (1) the new 3 deg$^2$ mosaic camera (called the Dark Energy Camera or DECam[7]) that will be installed on the Blanco 4m telescope at Cerro Tololo Inter-American Observatory (CTIO), and (2) the data management (DM) system[8] that will include astronomy modules, pipeline middleware, and archive, which will be released to the collaboration and the public. The DES project consists of a design and construction phase that began in October 2004 and continues through August 2009, and an operations and science phase that begins in September 2009 and extends through 2015, one year beyond the completion of observing (total of 525 nights over 5 years period). Below we describe the development of the DES data management system.

## 2. DATA VOLUME AND REQUIREMENTS

The DES will produce approximately 200 TB of raw image data in four bands over the course of the 5000 deg$^2$ survey, and these data will be used to produce deep, co-added images of the southern sky and precise photometric catalogs of the detected objects. The co-added images in four bands will require a total of about 70 TB of storage. Reduced and co-added images (32 bit) have associated variance images (32 bit) and bad pixel masks (16 bit), which are required for the object detection and measurement stages. Hence the final data volume for the reduced and co-added images will be five times larger than the raw image collection. In addition to these image collections, there is a catalog database, which will hold the photometry, astrometry and morphological information for the objects detected in the reduced and co-added images. With roughly 1 kB of information per object and approximately 1 billion objects to our depth over the survey region, we accumulate 1 TB of catalog data for each imaging layer that covers the entire region. Over the 5 year survey we expect to accumulate approximately 26 imaging layers, producing approximately 26 TB of catalog data. In addition, the catalog data from the yearly versions of the co-added images will accumulate at about 4 TB/yr. We estimate that the catalog database will approach 100 TB in size by the end of the survey. Therefore the total dataset will exceed 1PB. Because of the data volume, survey duration, and data quality requirements for our science, a highly automated, robust and accurate DM system is essential. Indeed, it appears that the requirements for automation and robustness in the DES DM system exceed those of any previous astronomy project.

The management of these raw and reduced data includes (1) immediate quality feedback on the mountain, (2) robust network transfer from CTIO to the archive center, (3) automated reduction to science-ready image and catalog products with quality assurance measures, and (4) long term archiving of the raw and reduced data products for the DES collaboration and the public. The archive center will be located at the National Center for Supercomputing Applications (NCSA) at the University of Illinois. Moreover, we will make our DES reduction pipelines and infrastructure available to the community users of DECam, who will enjoy about twice as many nights with this camera as the DES collaboration; one component of this service to the broader community will be a DECam Reduction Portal at NCSA, which will allow DECam users to employ our processing facilities to reduce their data.

## 3. CYBER-INFRASTRUCTURE FOR THE DARK ENERGY SURVEY

The design and development of the DES DM system is advancing through a close collaboration among computational astronomers and grid computing and database experts. Our guiding philosophy in designing the DES DM system is to build a system that takes advantage of existing computing infrastructure (both hardware and software) and incorporates a level of modularity and scalability that will allow us to upgrade our system to include future advances in astronomy modules, middleware and hardware. We designed our system to be easily altered to work for other instruments and projects, and in fact our development and testing is directly contributing to the data management project for the Blanco Cosmology Survey (BCS[9]) and the Large Synoptic Survey Telescope (LSST[10]).

The DECam will not be the first large mosaic camera; therefore, many of the data processing tools are already available. We are drawing from these available tools in building the DM system. However, the scale of the DES dataset requires a new level of automated processing and quality assurance. Developments in computing infrastructure, faster computing hardware and improved data handling experience within the astronomical community have informed our design of the DES DM system. This system will be used to process and archive the DES data in a way that takes advantage of existing computing infrastructure and ongoing development. We are adopting pipeline middleware that has been developed at NCSA and is being considered for the LSST data management. We have carefully examined the Sloan Digital Sky Survey (SDSS[11]) catalog archive in the design of our own catalog database, which will serve a dataset that is on the order of 100 TB. We will carry out our processing on publicly available high performance computing resources at NCSA and elsewhere, such as the TeraGrid[12] clusters. In addition, we are developing the system within a workstation environment, which ensures it will be available for the smaller scale DECam users. And finally, we are drawing upon existing, hardened image processing tools available within the astronomy community. Our approach of leveraging existing tools and expertise enables us to develop the DM system much faster and at far lower cost than some comparable astronomy projects that take the more traditional approach of designing all elements from scratch. Naturally, we hope that our successes will contribute to the development of future data management systems both within the astronomy community and beyond.

The DES DM system is a fully integrated archive and pipeline processing system, which is automated using a database repository that holds metadata describing our instrument, the survey, required calibration information and all the data acquired and processed. The DM system development is driven by the flow of the data from the time they are acquired until they are delivered as a science ready resource to the collaboration and the public. In Figure 1, we summarize the data flow for the DES. Data flow by network from the observatory to the main processing and archiving center at NCSA where the data are processed and results archived for later use. In addition to the archive and pipelines, the DM system is designed to provide a DECam reduction portal and Science Gateway. In the sections below we describe each of these components in greater detail.

### 3.1. DES Data Flow

DECam images and monitoring data move from the camera into a mountaintop, short-term archive. Here they are partially analyzed for immediate data quality feedback, and then transferred by network from CTIO to NCSA (with current bandwidth of 45Mbps and an additional DECam requirement, starting in 2009, of 36Mbps). At NCSA the data are ingested into the long-term archive, and pipeline processing begins. The process of image ingestion involves splitting the 1 GB multi-extension FITS (MEF) files into 62 individual CCD files that are ready for data parallel processing. Crosstalk corrections and initial adjustments to the world coordinate system (WCS) parameters and other header information are made at this time. A broad array of FITS header information is ingested into the DES database, enabling selection and retrieval of images at a later time through remote database queries. During the survey these raw data will arrive over a period of 18 hours starting at the beginning of the night; therefore, this ingestion and preprocessing is carried out serially. Figure 2 contains the DECam focal plane footprint, produced by displaying a single reduced DECam image with DS9[13].

After the data arrive, a master node generates and submits a series of jobs to the high performance computing (HPC) batch queues at NCSA (discussed below). Processed data are ingested from the HPC platforms directly into the DES archive. Integrated into this automated processing is the quality assurance testing that can affect the progress of the pipelines and trigger alerts for the DM team when intervention is required. Results of the full quality assurance of the image data will be stored in the DES archive database to enable quality filtering of the data. The processing of a night of raw data into reduced images and object catalogs will be a daily affair. The raw and reduced data will be archived at NCSA and made available for both the collaboration and the public.

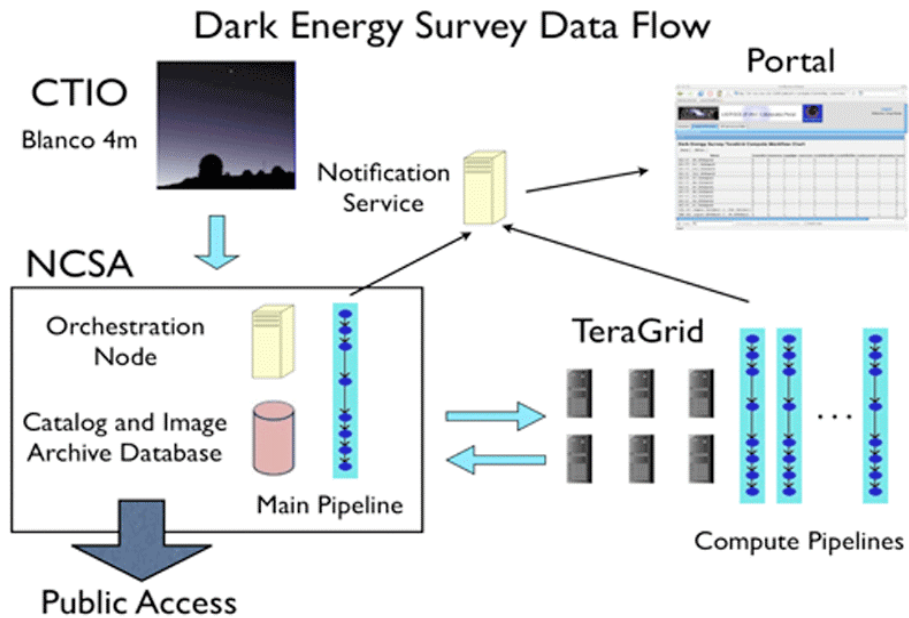

**Figure 1.** The data flow of the DES. Data move by network from CTIO to NCSA where automated processing and quality assurance begin. An orchestration node ingests the raw data into the DES archive, assembles the reduction pipeline and calibration data and then submits parallel jobs to TeraGrid for processing. The reduced data products are then ingested into the DES archive for calibration, scientific analysis and public access. Operators can monitor progress using a web portal that includes a notification service that reports pipeline events.

### 3.2. The DES Archive

The DES archive contains the raw and reduced images, catalog data, photometric and astrometric standards, monitoring data and the metadata required to analyze and understand the survey data. A relational database is used to contain the image and monitoring metadata together with the calibrated object catalogs. The DES archive is a distributed archive that encompasses the transport of the data from CTIO to the central archive at NCSA, as well as the satellite and mirror archive sites at collaborating DES institutions. All pieces of the DES archive have been designed and are being built by the DES DM team to suit the needs of our project. Additionally, we are collaborating with the NOAO Data Products Program (DPP) in the hope of employing an advanced version of the NOAO Science Archive (NSA) for public distribution of the DES image data; we also expect that other pieces of the software system NOAO has deployed to support the current instruments at CTIO, such as the data transport tools, will be available for use by the DES DM system.

#### 3.2.1. Image Archive

The image archive is fully integrated into the processing framework; a data access framework supports the simple data movement from archive to processing platforms and back, from spinning disk to mass storage and back, and among archive sites. The image data archive will be the central repository for all raw and reduced image products as well as monitoring data. It will include convenient interfaces to allow for access by the DM pipelines, the collaboration and the public. In addition, archive replication tools will be developed to enable all collaboration members to enjoy local access to the data. The image archive is under development by the DES DM team. We seek to build in compliance with the National Virtual Observatory (NVO) standards so that our data will be available through NVO portals, enabling joint analyses with other astronomical datasets. As mentioned above, we hope to employ the NSA in the public distribution of the raw and reduced image data products. Through collaborative development with the NSA project we leverage the dedicated efforts of DPP. As the DES comes to an end, we plan to work with NOAO to ensure the long term archiving and serving of these scientific resources to the community.

### 3.2.2. The Catalog Database

We have designed our catalog database to be fully integrated into the DES DM system and to stand as the primary science archive and analysis tool for the collaboration and the public. Our database schema are informed by the SDSS Skyserver, and our program of spiral development (detailed below) includes extensive testing of the database as both an integral part of the processing system and a powerful tool for scientific analysis and discovery. Our catalog database contains an image table with metadata (including quality flags) for each raw, reduced and co-added image that is catalogued. This facilitates easy linkage between objects and the image of origin, which enables filtering of objects using the characteristics of the images on which they were detected.

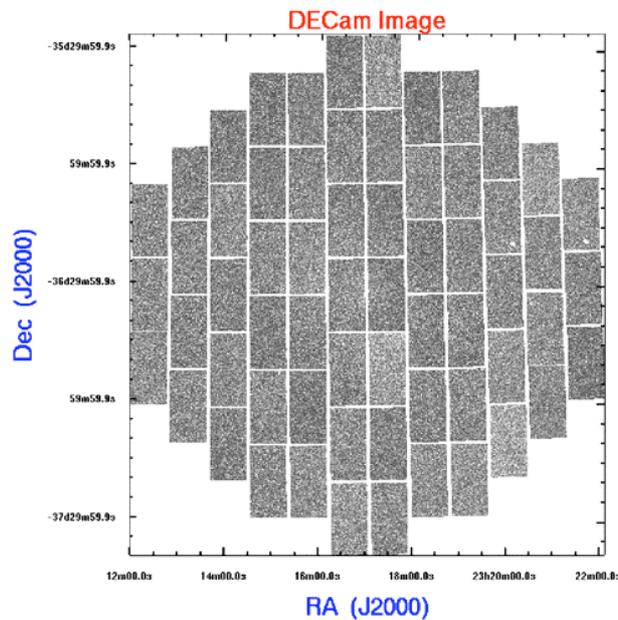

**Figure 2.** DECam's 62 individual CCDs are clearly visible in this reduced image. Most of the image reductions and cataloguing are carried out in data parallel fashion, speeding the process by a factor of 62 with no message passing overhead.

There are two novel aspects of our catalog database. First, we expect that image photometric zero-points will be refined over the course of the survey as better calibration data become available and as the global photometric calibration based on full information from objects that appear in multiple exposures becomes more accurate. Thus, we need to be able to incorporate this improved zero-point information into our catalogs while keeping an easily accessible history of previous zero-point measurements. To this end, our database includes a zero-point table that contains all accepted zero-points and detector color terms for each image in the survey, either from the latest zero-point or the history of the zero-points for a particular image. Thus, not only is the latest, most accurate photometry available for each object, but the object photometry from any previous instance can also be reconstructed. Second, we will not only catalog the reduced, single exposure images but also the co-added images. Our data model includes linkage from each single frame object to the corresponding co-added object of the same band. Much of our key science will be carried out using the cataloguing on the co-added images, but allowing for simple mapping from the co-added object table to the single frame object table will allow rapid access to light curves as well as multiple, independent morphological measurements for each object.

The sheer size and dual use of our database to support science queries and the pipeline processing system has led us to select an Oracle database for testing and development. We will use table partitioning to improve database response. This technique involves decomposing very large database tables into smaller pieces, so that queries are directed toward small partitions rather than entire tables. We will apply parallel processing to accelerate database performance, and we have selected a flexible, multi-node platform for development. Having multiple nodes that are all connected to the database through a shared file system allows for some computing resources to be reserved for ingestion and querying by the DES pipelines and other resources to be allocated for science queries by the collaboration and the public.

**3.3. Processing Pipelines**

DES pipelines make use of a general framework that is designed to promote automation and robustness against failures. This framework features a three-layer architecture: the Pipeline Orchestration Layer, the Pipeline Execution and Monitoring Layer and the Application Layer. The pipelines themselves are astronomy processing modules organized using modern workflow management tools that are being developed at NCSA.

At the top, the *Pipeline Orchestration Layer* is responsible for determining what processing needs to be done and for executing the appropriate pipelines. The DES data processing is organized and controlled by a pipeline orchestration node. In general, the orchestration layer is triggered by the arrival of new data into the archive; however, operators have both command line and web interface controls. The orchestration layer assembles a pipeline by drawing modules from a pipeline library and generates a pipeline configuration on-the-fly. This layer is also responsible for assembling the data that are needed by the pipeline and staging them on the appropriate platforms by sending queries to the image database, and using the data access framework to pre-position these data to the file systems of the processing nodes. After the pipeline has been assembled and the data have been staged, the orchestration layer executes the pipeline on a remote computational platform at NCSA (or elsewhere) and monitors its progress using a notification service. The processing strategy is to generate 62 parallel jobs (one for each DECam CCD; see Figure 2 for the layout of the CCDs) and submit to NCSA TeraGrid clusters that carry out the basic image reduction to the catalog level. When the pipeline processing is complete, the orchestration layer ensures the return of the resulting data products to the storage cluster for ingestion into the archive. The availability of reduced images and new catalogs typically triggers additional processing for calibration, co-addition and another round of cataloguing on the co-added images.

The second layer is the *Pipeline Execution and Monitoring Layer* that actually launches the pipelines on remote computational platforms. This will be powered primarily by the Condor job submission and management system, accessed via Condor-G, the interoperability interface between Condor and Globus grid middleware. Condor-G provides the system with the portability that will allow us to harness computational platforms beyond those at NCSA. We plan to leverage a capability being developed by the National Middleware Initiative and Open Science Grid that will make it straight-forward to deploy our pipelines on other HPC platforms at partner sites. This will make it possible to execute pipelines on a broader array of computational resources beyond those at NCSA. We expect this capability to be especially important during full reprocessing and co-adding the reduced data. The pipeline monitoring occurs through a built-in event channel that allows quality assurance test results within each pipeline stage to be reported to the notification service where it is posted to the web for operator viewing and perhaps used to trigger alerts.

At the bottom is the *Application Layer*. Here application modules are stitched together to form a single data-parallel pipeline. It is also well suited for stitching together modules, particularly when the modules are self-contained. A convenient wrapper, i.e. a middleware, will be used for defining configurable application modules that can be organized into a module library and assembled into pipelines (by the orchestration layer) in different ways. In particular, as our software evolves and matures, we expect to create new versions of various modules. A module tracking capability is built into our DES database, which allows us to keep track of which data have been processed with which version of each module. This in turn will enable efficient and intelligent reprocessing of data as we incorporate improved modules into the library.

The pipeline framework is an area where our project has important strengths. Our approach allows us to fully leverage leading grid computing software and development tools. Going this route rather than writing our own special purpose grid computing tools allows us to (1) concentrate our efforts on the more specialized astronomy components of the DM system, (2) benefit from the many strengths of these leading components and the future improvements to these tools, and (3) create a more portable system that can be deployed on a range of platforms rather than being "hard wired" to work on a specific cluster available to only one project. This portability will be especially important during the computationally taxing co-addition and reprocessing, when we benefit from engaging high-performance computing platforms across the country. As our approach to modularity, scalability and portability proves its effectiveness in real world settings, we expect the system to generate broad interest within the astronomy community and perhaps even beyond.

### 3.3.1. Pipeline Middleware

Currently, we are using the Open Grid Runtime Engine (OGRE), a stable and mature grid workflow management tool being developed at NCSA, as the primary middleware for the three layers discussed previously. The OGRE project is led by A. Rossi and S. Hampton within the Middleware Division of the Integrated Cyberservices directorate at NCSA. OGRE is a Java and XML based workflow language that was developed under the auspices of projects such as the Alliance Portal Expedition, the Open Grid Computing Environments (OGCE[14]) Collaboratory and the Linked Environments for Atmospheric Discovery (LEAD) ITR, with J. Alameda leading these NCSA efforts. The XML format that constitutes an OGRE script provides a convenient wrapper for defining configurable application modules, and OGRE's "include" functionality make it straightforward to assemble the pipelines in different ways. The OGRE "scripts" will also contain the intelligence for determining (primarily via database queries) what needs processing and what modules to use with, e.g., the looping and conditional execution capabilities. Beyond the local flow of control, OGRE supports numerous operations typically required by grid aware applications; it interfaces to NCSA's Trebuchet file transfer library to enable GridFTP-based file transfers and remote directory listing. OGRE supports the execution of underlying Unix threads, and hence OGRE can wrap arbitrary application codes (Condor-G clients, astronomy module binaries, database calls) within a workflow. Another important feature of OGRE that we will use in the application layer is the event channel, which provides a way to send status and error messages back up to a notification service. The notification service can collect messages from all running pipelines and display them for administrators as well making them available to the orchestration layer. This will be key for providing robustness and minimizing human intervention. In the exploration and prototyping of grid middleware solutions for pipeline processing and data management, the DES DM team has collaborated with the LSST DM team at the University of Illinois.

### 3.3.2. Astronomy Modules

The astronomy modules are the basic building blocks of the pipelines. In our system we have built the following modules: (1) the single frame module, which does the typical image reductions (overscan, trim, bias subtract, flat field, pupil ghost, illumination correction, etc), (2) the astrometric solution module, which refines the WCS information in the image headers, (3) the cataloguing modules, which do basic object detection as well as more sophisticated morphological measurements, (4) the photometric standards module, which is used to calculate photometric solutions from standard star observations, (5) the photometric calibration module, which applies the photometric solutions to the images and catalogued objects, (6) the co-add module, which does the image remapping and combining to produce the deeper, more complete co-added image collection, and (7) the global photometric calibration module, which will use the overlapping observations as well as the available zero-point information to calculate a uniform photometric solution over the full survey area.

The module accuracy must meet the data quality requirements that derive from the scientific analyses. In addition, these modules must be modular and easy to port/install on our computing platforms. We are using existing, hardened modules as a starting point whenever possible. Through the process of testing and validation we will determine which modules require further development to meet the data quality and performance requirements. For the first data challenge (DC1; further described in Section 4), we employed the following publicly available astronomy modules: SExtractor[15], SWarp[15], WCSTools[16] and ANNz[17]. In addition, during the development leading to DC1, we explored IRAF (Image Reduction and Analysis Facilities[18]), the Pan-STARRs related package *mana*[19], and a Perl-based MOSAIC pipeline system that employs elements of both IRAF and *mana*. Our experience with these packages led us to write our own basic image correction codes for the single frame module, which are more portable and better meet the needs of the DES. In particular, our reduction code follows individual CCDs (with two amplifiers), and calculates inverse variance images and bad pixel masks that encode the origin of masked pixels. These variance images and bad pixel masks, along with the reduced images themselves, are critical inputs for the co-addition module.

Our iterative development strategy involves cycles of testing and development that will drive us to extend community codes and the codes we are developing from the ground up. For example, we plan to develop an astrometric calibration module that uses the full focal plane dataset along with a well-characterized model of the focal plane distortions. This full focal plane solution should provide improved accuracy and can be implemented through queries to the catalog database with no need for the image data. Similarly, we are currently studying the way SExtractor carries out deblending and measures magnitudes within the context of varied seeing (in different bands, on different nights, over the focal plane); meeting our photometric accuracy requirements will likely involve additional development.

### 3.4. Science Gateway and DECam Reduction Portal

The archived data at NCSA will be made available to both the collaboration and the public throughout the survey. Raw and reduced images will be released one year from the date they are acquired. Higher level data products will be released to the public in two stages: once in the middle of our survey and then again one year after the completion of the survey. Therefore, we plan to develop a DES Science Gateway to facilitate the scientific analysis of DES archival data using science modules contributed by the collaboration and by the public. Because the DECam will be used by non-DES observers, we will also provide access to the DM system code and will develop a DECam Reduction Portal to aid in DECam data reduction for these non-DES users.

The DES Science Gateway will enable users to execute compute jobs on HPC resources that carry out scientific analyses of the DES catalog data using selected science modules. For example, one might want to estimate galaxy cluster photometric redshifts using an input list of cluster positions together with the DES catalog data, or one might wish to construct a weak lensing map around a particular position using DES catalog data. These examples are modules that the DES collaboration will develop to support the key dark energy science projects, and we will make these available through our Science Gateway. The possibilities for other scientific tools are endless; our plan is to construct a framework for this Gateway and then allow the collaboration and the public to contribute modules. These contributed modules would be extensively tested with the help of the author(s) before being made available for general Gateway use.

An important commitment that we have made to NOAO is to release the DM system that we are developing for the DES to other users of DECam. Therefore, we plan to develop what we term a DECam Reduction Portal. This portal will essentially be a self-service web interface to our DES reduction system, where non-DES users could trigger the reduction of their DECam data. Moreover, the flexible and scaleable nature of our DM system will make it straightforward to work with users to configure their own pipeline. For example, they could request an alternate cataloguing module for their analysis or choose to use our cataloguing module with a non-DES parameter set. The DECam Reduction Portal is similar to the pipeline management tool that we are developing for our own use in the DES DM system, so creating the DECam Reduction Portal will require adding public access with an appropriate security layer and providing support. Note that the distributed nature of our pipeline processing framework naturally allows for orchestration and data processing to be in separate locations, and this will simplify the inclusion of additional computing resources beyond those we exercise daily in the DES reductions.

### 4. TESTING AND VALIDATION

We have adopted a spiral or iterative development strategy for the DM system. That is, we have cycles of development followed by periods of exhaustive testing that we call Data Challenges (DC). During the 2004-2009 development period, there are four Data Challenges beginning in October of each year from 2005 through 2008. Our Data Challenges will be followed each February by an examination of the performance of our system with respect to the DM requirements, which will allow external experts to determine whether we are making appropriate progress and to recommend changes to our development strategy. The evaluation of the current system each year is then used to refine the development plan for the next cycle. The spiral development strategy offers significant risk mitigation for our project, because we have a working DM system in place very early in the process, which we can use in a process of testing and discovery. We then focus on improving components and adding functionality. Table 1 contains an overview of our plans for the Data Challenges.

The Data Challenges involve using our DM system to reduce simulated DECam data, which is made increasingly realistic over time. In addition, we are testing our pipeline by reducing MOSAIC II (currently installed on the CTIO Blanco 4m telescope) camera data obtained as part of the ongoing BCS[9]. Over time there is a ramping up of the volume of imaging data until in the third season (Oct 2007) we carry out a stress test that includes an entire simulated season of observing (i.e. 105 nights). The simulated images are produced at Fermilab by the DES simulation team, and they include (or will include) a wide range of realistic effects: (1) seeing variations that reflect the measured distribution on the Blanco + MOSAIC II, (2) field distortions and spot sizes calculated using the design of the new corrector, (3) bad pixels, bad columns and edge effects consistent with the specifications of the DECam detectors, (4) cosmic rays, (5) pupil ghost at the 5% level, consistent with expectations given the optical system, (6) read noise and gain variations consistent with our expectation, (7) sky brightness variations due to lunar phase, (8) pointing errors and differential refraction across the field, (9) saturation and trails from bright stars, and (10) modeling of non-photometric nights and other effects as needed. The actual input catalog data include stars from the USNO-B catalog together with galaxies sampled from large-scale structure simulations. These galaxies are appropriately clustered and sheared in a manner consistent with the simulated matter distribution along the line of sight, and they have realistic colors and morphologies that enable direct tests of our photometric redshift and shapelet cataloguing modules. In addition to the simulated images, the truth tables (tables containing the input positions, brightnesses and morphological parameters for each object) are all ingested into our DES database. The catalogs produced from the reduced images are then compared to the truth tables, providing information about accuracy of our photometry, astrometry, photometric redshifts and galaxy morphologies. Finally, these catalogs undergo a scientific analysis for an end to end test of our ability to find galaxy clusters and estimate redshifts, recover the clustering of the galaxies, recover the weak lensing shear distribution and measure light curves of supernovae Ia.

Table 1: Dark Energy Survey Data Management Testing Schedule

| Date | Dataset | Development Goals |
|---|---|---|
| DC0<br><br>Dec 04-<br>Mar 05 | 100 deg$^2$<br>One layer deep in each band<br>(100 GB raw) | Convert existing pipelines (Rest et al., Valdes et al.)<br>Test on workstations with simulated DECam data |
| DC1<br><br>Oct 05-<br>Jan 06 | 100 deg$^2$<br>Five layers deep in each band<br>(700 GB raw, 5 TB reduced) | Test DM system v1 on TeraGrid<br>Image reduction to catalog level (overscan, bias, flat, pupil ghost, astrometry)<br>Image and catalog data ingestion<br>Full database-query driven operations |
| DC2<br><br>Oct 06-<br>Jan 09 | 300 deg$^2$<br>Five layers deep in each band<br>with additional features<br>(2.1 TB raw, 15 TB reduced)<br><br>50 deg$^2$ *griz* imaging from<br>Blanco Cosmology Survey<br>(Blanco 4m + MOSAIC II) | Test DM system v2 on TeraGrid<br>Retest all the above elements, plus: global photometric calibration, co-add with 5 layers in each band, automated quality assurance<br>Pipeline error tolerance<br>Archive replication<br>Data transfer from remote site<br>Test with real data (BCS images) |
| DC3<br><br>Oct 07-<br>Jan 09 | One observing season of data<br>(36 TB raw, 250 TB reduced) | Stress test DM system v3 on TeraGrid<br>Deploy and test at designated collaboration sites<br>Science queries on database through web interface |
| DC4<br><br>Oct 08-<br>Jan 09 | Same data as before with<br>additional defects in images | Test operations version of DES DM system (v4) on TeraGrid and at collaboration sites<br>Test DECam Reduction Portal<br>Test Science Gateway<br>Exercise mountain operations components |

Early on in spring 2005, we had a preliminary data challenge (DC0), when we altered two existing pipelines to process simulated DECam data into object catalogs. One of these pipelines has been used to reduce data from the Blanco 4m MOSAIC II camera for several years. The other pipeline is under development by the NOAO IRAF team for the MOSAIC cameras on CTIO and KPNO (Kitt Peak National Observatory). Both of these pipelines rely on the venerable IRAF astronomy reduction package, and so our tests gave us important insights into the challenges of deploying IRAF within a modern, grid computing environment. We used this initial Data Challenge as a launching point into completing the design and beginning the development of the DES DM system. More recently we finished Data Challenge 1 (DC1), which involved the deployment and testing of the first version of our DES data management system. We describe these activities in more detail below.

As outlined in Table 1, our goals in DC1 were to deploy and test a working version of the DM system (described in detail in Section 3). In the end we succeeded in testing the DM system without the co-addition stage. The hardware configuration used for DC1 included the following Linux systems: (1) a 4 node storage cluster with a 6 TB parallel (lustre) file system, (2) a quad CPU database platform running Oracle, (3) an orchestration node, and (4) the compute nodes of the NCSA TeraGrid Mercury cluster[**]. Data were transferred from Fermilab onto our storage cluster. The 700 GB of raw images were organized into five "nights" of observing with associated bias and flat frames available for each. Each night was photometric with different extinction coefficients and time and band variable seeing sampled from the measured distribution at CTIO. The testing proceeded through each stage of the processing. First, we split the images to enable data parallel processing and ingested the image metadata into our DES database. Interactions with the database platform were allowed from both the storage cluster and the orchestration node. We then reduced the data one night at a time through 62 independent compute jobs.

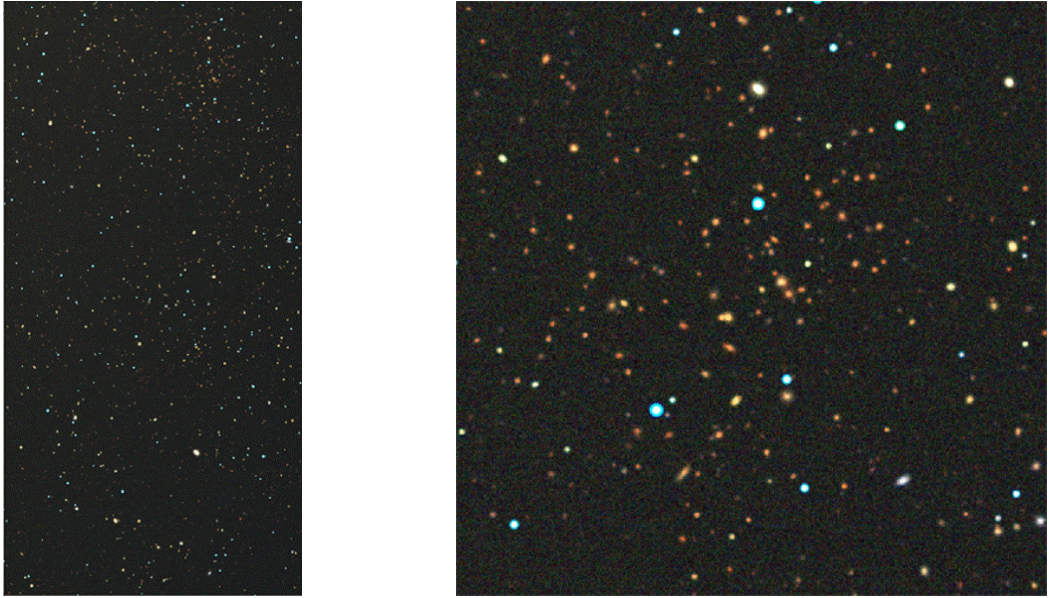

**Figure 3.** (a) A single CCD, pseudo-color image constructed from reduced DC1 *g*, *r* and *i* band images. (b) An enlarged region of the reduced image showing the galaxy and stellar distribution in one part of the simulated images, which were created from catalog level simulations by the DES simulation team at Fermilab.

For each CCD, the nightly reductions begin with a query to the DES database to produce lists of all available raw science and calibration images. We then use the initial WCS information for each of those science images to query the USNO-B table to create a single astrometric standards list for that job. Following this the raw data, reduction pipelines, image lists and calibration data are staged to the Mercury cluster parallel file system (accessible from any Mercury node) with a unique job identifier in the directory tree. The processing job requesting a single node is then submitted to the Mercury batch queue. When this job is launched on one of the Mercury nodes, it finds the appropriate data collection in the shared file system and begins reductions. The first stage is the single frame reduction stage, which overscans and combines all bias images into a bias calibration image and then overscans, bias corrects and combines each band specific set of flat field images into flat calibration images. Bad pixel masks are provided as initial calibration inputs, and variance images are produced for each of the calibration products. The calibration products are then used to overscan, bias subtract, flat field correct and interpolate over bad columns in all the science images. Reduced science images include a bad pixel mask with saturated and interpolated pixels marked, and a variance image in a single MEF image. Because there are two amplifiers per CCD, the reduction code must deal with the overscan correction of each amplifier independently. The single frame reduction codes are implemented in the *C* programming language by the DM team. The following stage is refinement of the WCS astrometric parameters in each science image header. In DC1 this involves rapid cataloguing using SExtractor followed by catalog matching and refinement using WCSTools (*imwcs*). After that, each image is catalogued with SExtractor, using the variance image as a weight image and the bad pixel mask as a flag image. SExtractor runs to produce a binary FITS catalog, an object region image and a sky subtracted object image. These outputs are combined into a single MEF file. A second cataloguing stage developed at Fermilab then calculates adaptive moment coefficients for using the output data from the first stage.

---

[**] The NCSA Mercury cluster is a Linux cluster that contains 887 nodes, with 1/3 and 2/3 of them running on dual 1.3 and 1.5 GHz Intel Itanium 2 processors, respectively.

In Figure 3 we show some reduction results from our automated pipelines. Reduced image data are five times larger than the original input science image, and in combination with the cataloguing results the total reduced dataset is approximately 7.5 times larger than the input data. After a single night is processed by 62 simultaneous compute jobs, the reduced data are then retrieved from the compute nodes and placed onto the storage cluster using OGRE modules that employ a multi-streamed GridFTP process. Images and catalog data are ingested into the DES archive for further processing. In DC1 these additional stages were simply for photometric calibration and testing. For each band and night, we trigger a database level photometric standards module that retrieves all objects from standard star observations and their associated matching objects in our photometric calibration catalog. It then calculates a photometric solution for each CCD and stores that solution in another DES database table. A photometric calibration module selects solutions for each CCD, night and band and calculates the zero-point for all appropriate science images and uses those zero-points to update the photometric zero-point for all detected objects in each image. Through these steps we succeeded in reducing and calibrating 100 $deg^2$ of "sky" with five layers of imaging in each of the *griz* bands (~100,000 individual CCD images). DC1 efforts catalogued and calibrated 50 million objects. We tested the results by visual inspection of some of the images and by comparing the measured photometry and astrometry in the DES database against the truth tables used to construct the simulated data. In these tests we actually discovered errors in the initial pointing information placed in the headers of some of the simulated images. After correcting the headers of the simulated data, tests indicate reasonably good astrometric (0.25 arc-second RMS) and photometric accuracy (<<0.01 magnitude zero-point offset).

## 5. CONCLUSIONS

The DES project is designed to study the cosmic acceleration or, perhaps equivalently, the nature of dark energy using four complementary and independent techniques. The four techniques that will be employed by the DES are considered the key techniques for studying the nature of dark energy; these are: (1) a galaxy cluster survey in collaboration with the South Pole Telescope mm-wave mapping experiment, (2) weak lensing measurement of the cosmic shear, (3) galaxy angular power spectrum measurements, and (4) supernovae Ia distances. An intense program of scientific development is underway to characterize and control systematics in these four methods so that, in combination, they can be used to differentiate between models of dark energy and an underlying flaw in General Relativity at large separation. NOAO has agreed to grant the DES collaboration 525 nights on the Blanco 4m telescope on CTIO over the 5 years of the survey in return for delivering (1) the 3 $deg^2$ DECam mosaic camera, and (2) the data management system, and (3) the archive of raw and reduced DES survey data. The survey will generate 200 TB of raw data. When the DES project ends, the total data volume from the raw and reduced data will exceed 1 PB. We have developed the DES DM system to be flexible, scalable, robust and highly automated so that we can efficiently and accurately manage this dataset.

Our DM system is built to leverage the high performance computing resources of the TeraGrid and ongoing development in astronomical software and computational middleware. This DM system consists of integrated processing pipelines and an archive, which includes a DES database. The DES database is used for storing and analyzing catalog data as well as for storing and retrieving all the image metadata, calibration standards, and survey metadata that are needed to enable automated operations. Ultimately, our DES DM system will include (1) a Science Gateway to support compute and data intensive analyses on publicly available HPC resources and (2) a DECam reduction portal that will simplify processing of DECam data acquired by non-DES users. Our DM system is built using a workflow management tool under development at NCSA, which wraps astronomy processing modules that build upon community hardened codes. As detailed in previous sections, the reduction process includes (1) image data reduction to the catalog level, (2) co-adding, (3) photometric and astrometric calibration and (4) production of final catalogs. Our development plan and strategy includes a series of Data Challenges that mark the continued refinement and testing of our DM system using increasingly realistic simulated data along with large quantities of data from the existing MOSAIC II camera at CTIO.

In January 2006 we completed the deployment and testing of the first version of our DM system. As detailed above, we have a fully integrated archive and pipeline processing system in place that we have successfully used to reduce 700 GB of raw data into 5 TB of reduced data products and to catalog and calibrate 50 million objects. Currently, we are developing our DM system for the next step of testing during DC2, beginning in October 2006. In closing, we note that some previous projects (and even some current large scale optical imaging surveys) have tended to focus on (re)developing middleware from the ground up rather than leveraging the enormous investments already made in the infrastructure required for large scale data management. Reinventing the wheel is expensive and it wastes time; our strategy of leveraging existing software and NCSA expertise is largely responsible for the rapid development, deployment and successful testing of the first version of the DM system. We hope to build upon these early successes as we complete the development of the DES DM system.

## ACKNOWLEDGEMENTS

We acknowledge helpful conversation with many members of the DES team, and we extend our thanks especially to John Peoples, Brenna Flaugher, Wyatt Merritt, Chris Miller and Chris Stoughton. We acknowledge critical seed funding and hardware support from the Astronomy Department, the College of Liberal Arts and Sciences, the Office of the Vice Chancellor for Research and the NCSA Director at the University of Illinois. We wish to thank the TeraGrid project for supporting the DES DM project through the allocation AST050027T. The TeraGrid project is funded by the National Science Foundation.